\documentclass[numberedappendix,iop]{emulateapj}

\usepackage{graphicx}
\usepackage{natbib}
\usepackage{ulem}
\usepackage{txfonts}

\slugcomment{Submitted to the Astrophysical Journal}

\shorttitle{}
\shortauthors{Mart\' inez Gonz\'alez et al.}

\begin{document}

\title{On the magnetism and dynamics of prominence legs hosting tornadoes}

\author{M. J. Mart\'\i nez Gonz\'alez\altaffilmark{1,2}, A. Asensio Ramos\altaffilmark{1,2}, I. Arregui\altaffilmark{1,2}, M. Collados\altaffilmark{1,2}, C. Beck\altaffilmark{3}, J. de la Cruz Rodr\' iguez\altaffilmark{4}}
\affil{Instituto de Astrof\'\i sica de Canarias, V\'\i a L\'actea s/n, E-38205 La Laguna, Tenerife, Spain}
\altaffiltext{2}{Dept. Astrof\' isica, Universidad de La Laguna, E-38206, La Laguna, Tenerife, Spain}
\altaffiltext{3}{National Solar Observatory, Sacramento Peak P.O. Box 62, Sunspot, NM 88349, USA}
\altaffiltext{4}{Institute for Solar Physics, Department of Astronomy, Stockholm University, Albanova University Center, SE-10691 Stockholm, Sweden}

\begin{abstract}

Solar tornadoes are dark vertical filamentary structures observed in the extreme ultraviolet associated with prominence legs and filament barbs.
Their true nature and relationship to prominences requires understanding their magnetic structure and dynamic properties. 
Recently, a controversy has arisen: is the magnetic field organized forming vertical, helical structures or is it dominantly horizontal? 
And concerning their dynamics, are tornadoes really rotating or is it just a visual illusion?
Here, we analyze four consecutive spectro-polarimetric scans of a prominence hosting 
tornadoes on its legs which help us shed some light on their magnetic and dynamical properties. 
We show that the magnetic field is very smooth in all the prominence, probably an intrinsic property of the coronal field. 
The prominence legs have vertical helical fields that show slow temporal variation probably related to the motion of the fibrils. 
Concerning the dynamics, we argue that 1) if rotation exists, it is intermittent, lasting no more than one hour, and 2) the observed velocity pattern is also consistent with an oscillatory velocity pattern
(waves).

\end{abstract}


\maketitle

\section{Introduction}

The term solar tornado was first introduced by \cite{pettit_32} to name a specific kind of solar prominences that appeared like ``vertical spirals or tightly twisted ropes". Recently, the high spatial resolution, cadence and continuity of the coronal data provided by the Atmospheric Imaging Assembly \citep[AIA;][]{aia} 
onboard the Solar Dynamics Observatory \citep[SDO;][]{sdo} has re-awakened the interest in the study of solar tornadoes. 
Solar tornadoes are nowadays associated with vertical funnel-shaped dark structures seen in the coronal 
Fe\,{\sc ix} line at 17.1 nm. These structures are found to be hosted in the legs of some quiescent prominences \citep[e.g.,][]{su_12,wedemeyer_13} and seem to be directly linked to the evolution and final fate of the prominence \citep{li_12,mghebrishvili_15}.

The magnetic and dynamic properties of solar tornadoes are still under debate. Concerning the magnetic configuration, \cite{yo_15} found that they harbor helical fields connecting the main body of the prominence with the 
underlying solar surface, in agreement with previous claims of vertical fields in filament barbs \citep[or prominence legs;][]{zirker_98}. On the contrary, \cite{brigitte_15} inferred magnetic fields that are almost parallel to the surface, 
in agreement with the magnetic field measured in a photospheric barb endpoint \citep{arturo_06}. 

Concerning the dynamics, it is unclear whether there is a real \citep[e.g.,][]{su_12} or only apparent rotation \citep[e.g.,][]{panasenco_14} of the structure. An eventual rotation together with plasma flows could support the 
heavy cool plasma in a hypothetically vertical field of a solar tornado \citep{luna_15}. 
\cite{su_12} reported on the oscillatory pattern in the plane-of-the-sky motions of solar tornadoes as seen 
in SDO/AIA 17.1 nm, claiming that 
these structures are rotating with periods of about 50 min. Subsequent works \citep{david_12,wedemeyer_13, levens_15} confirmed the 
rotation scenario by measuring oppositely-directed Doppler shifts at both sides of prominence legs. \cite{su_14} showed evidence of a sustained rotation in tornadoes over a course of 3 h. \cite{panasenco_14} proposed that the rotation of solar tornadoes as seen in SDO/AIA 17.1 nm is only apparent and could be explained by waves. Recently, \cite{mghebrishvili_15} explored both the rotation and the wave scenario and found both to be compatible with SDO/AIA 17.1 nm data of solar tornadoes. They also proposed that, most likely, both rotation and waves are at work, identifying the oscillation as a magnetohydrodynamic kink mode. 

Here, we present a complete set of spectro-polarimetric observations of a quiescent prominence. The dynamical properties of its legs (they show a sinusoidal motion in the plane of the sky), and its helical appearence make us identify them as the solar tornadoes of \cite{pettit_32} and \cite{su_12}. Spectro-polarimetry allows us to shed some light on the magnetic and dynamical properties of prominence legs that host solar tornadoes.

\begin{figure*}[!ht]
\includegraphics[width=\textwidth, bb=35 19 447 81]{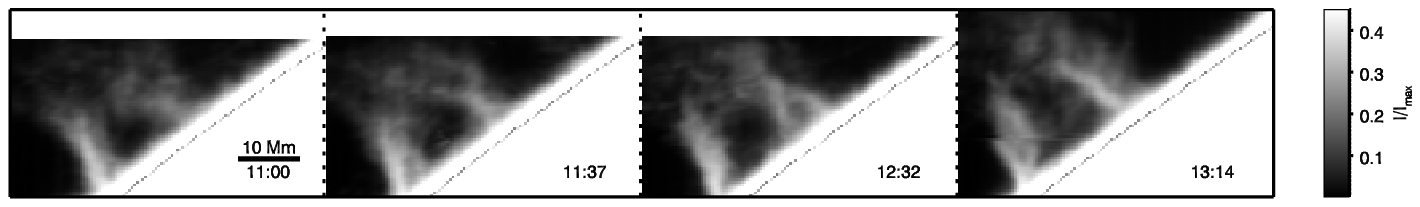}
\includegraphics[width=\textwidth, bb=35 19 447 81]{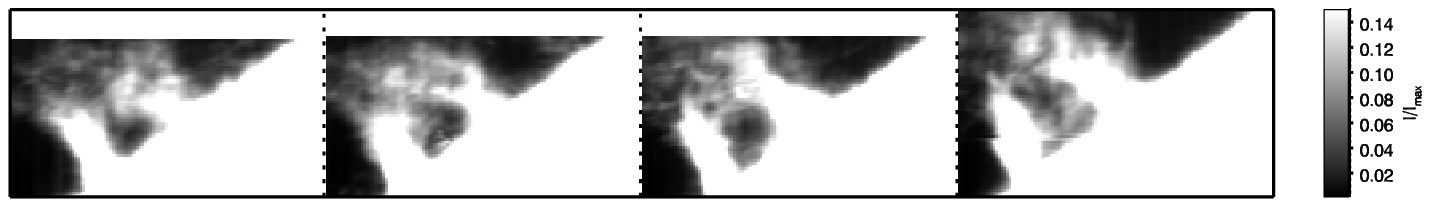}
\includegraphics[width=\textwidth, bb=35 19 447 81]{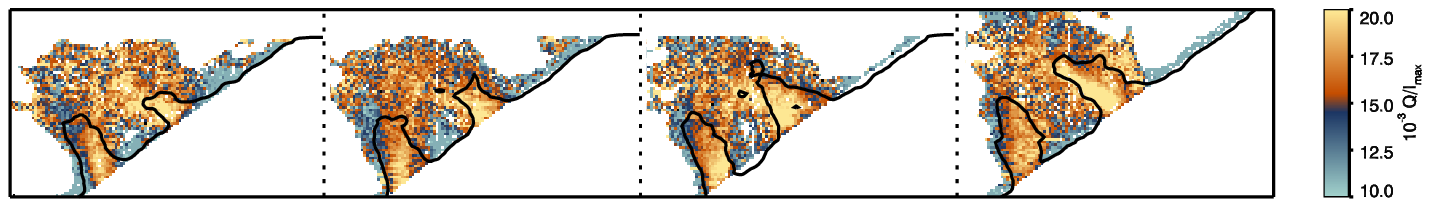}
\includegraphics[width=\textwidth, bb=35 19 447 81]{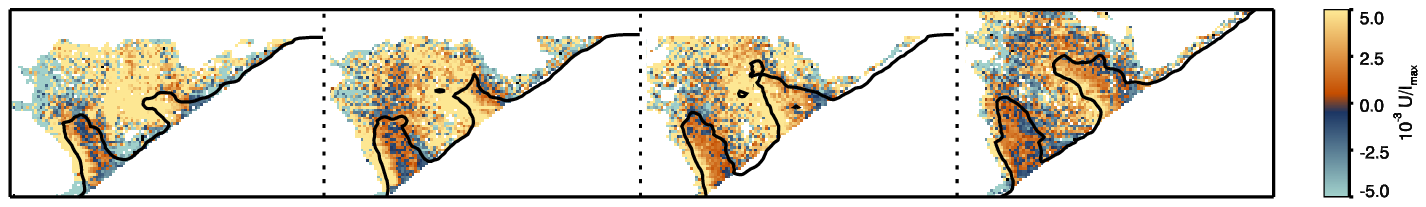}
\includegraphics[width=\textwidth, bb=35 19 447 81]{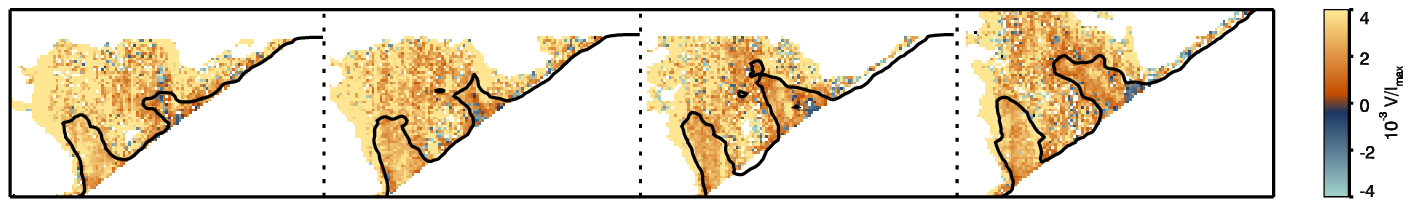}
\caption{Full Stokes spectro-polarimetry of a prominence. The topmost panels display the 
intensity maps at the core of the He\,{\sc i} line with two different saturation levels. The rest 
of the panels display the polarization signals relative to the maximum intensity at each pixel. The 
superimposed contours enclose He\,{\sc i} line-core intensities larger than 5$\times 10^3$ I$_\mathrm{max}$, 
containing most of the plasma of the prominence legs. The slit was oriented parallel to the horizontal 
direction and the scan was performed from the bottom to the top. The times are given in UT.}
\label{intensidad_he}
\end{figure*}

\begin{figure*}[!ht]
\includegraphics[width=0.2\textwidth]{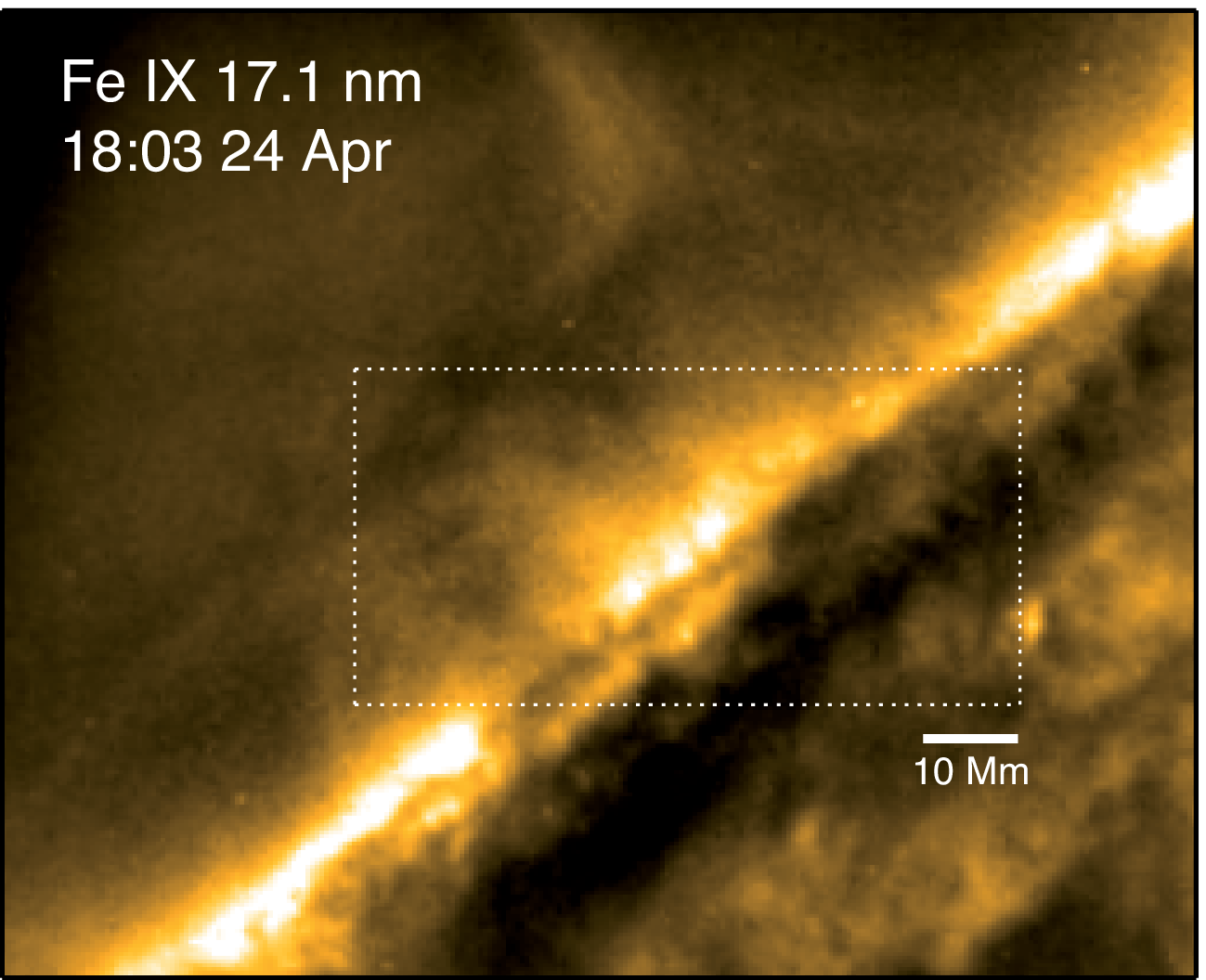}
\includegraphics[width=0.8\textwidth]{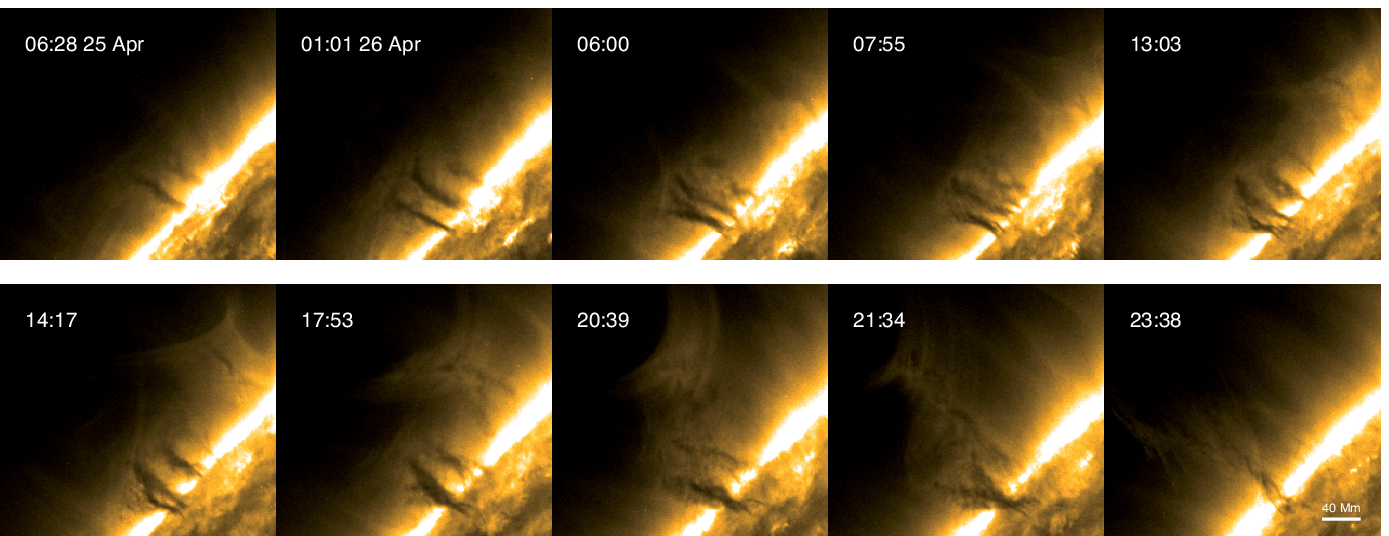}\\
\includegraphics[width=0.2\textwidth]{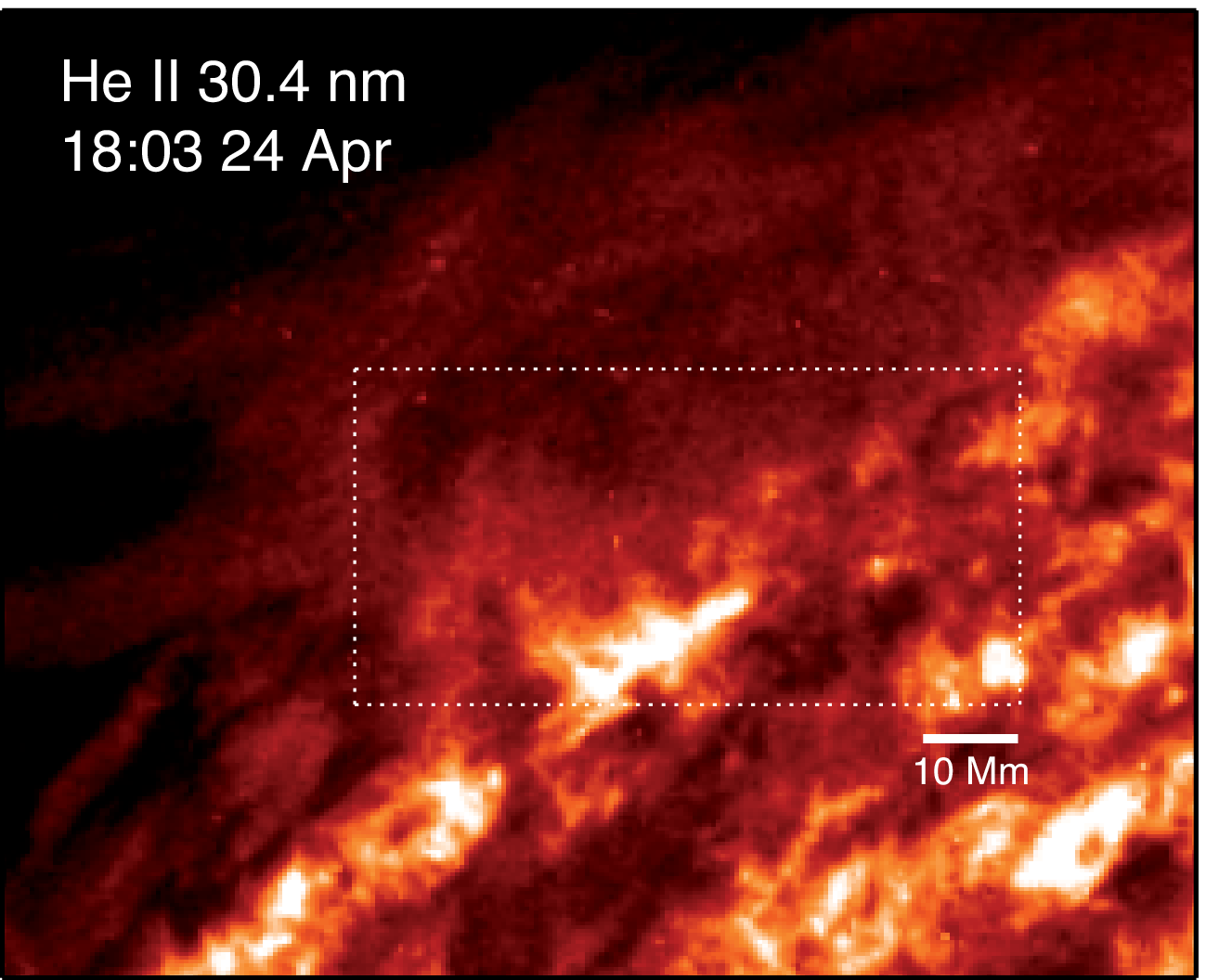}
\includegraphics[width=0.8\textwidth]{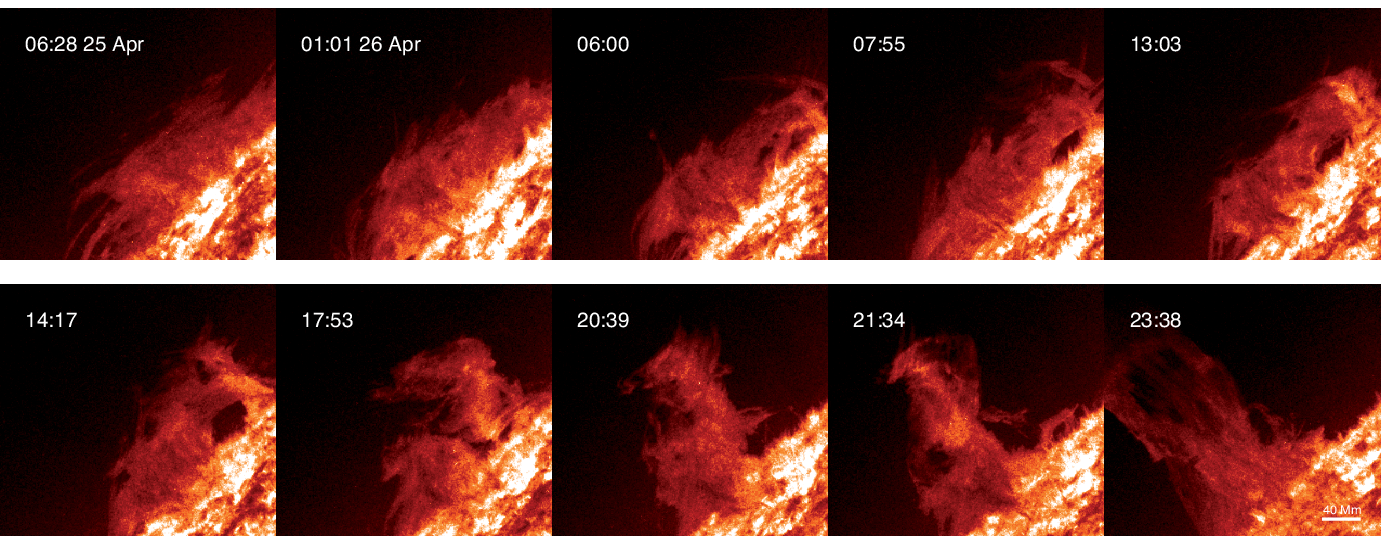}
\caption{Evolution of the target prominence as seen in the Fe\,{\sc ix} 17.1 nm (top panels) and the He\,{\sc ii} 30.4 nm 
coronal lines. At the time of our spectro-polarimetric observation (around noon April 24, 2012) the prominence was behind the limb and hence it was 
hidden by the glow of the coronal lines. An identification (see the leftmost panels) was possible 5 h after the spectro-polarimetric observations. The times are given in UT.}
\label{images_aia}
\end{figure*}

\section{Observations and data analysis}

\subsection{Spectropolarimetric observations}

We performed spectro-polarimetric observations of the chromospheric He\,{\sc i} line near \hbox{1083 nm} 
of a quiescent prominence. These observations were performed with the TIP-II instrument \citep{tipII} at the German Vacuum Tower Telescope on the Observatorio del Teide. This quiescent prominence appeared at the East limb (latitude of 42$^\circ$S) on April 24th 2012. We performed four scans covering an area of $\sim 33-42'' \times 72''$ that contained most of the prominence material. 
The integration time was $\sim 30$ s at each slit position. The first observation started at 11:00 UT, the second one at 
11:37 UT, the third one at 12:32 UT, and the last one started at 13:14 UT. On average, the total time spent for 
one map was about half an hour. The reason for the time gap of about one hour between the second and the third map was a flatfield and some aborted observations. The atmospheric conditions were excellent and the image was so stable that the adaptive optics (AO) system worked flawlessly during the whole observation period, allowing us to achieve a spatial resolution close to the diffraction limit of the telescope at 1 $\mu$m ($\sim \sim 0.6''$). The pixel size along the slit was 0.16$''$ while the scan step was set to 0.6$''$. This step size allowed us to scan the prominence in a reasonable 
time). In order to increase the signal to noise of the observations, we averaged two pixels along the 
slit. The noise in polarization is, on average, $7\times 10^{-4}$ in terms of the maximum intensity 
(I$_\mathrm{max}$). 

We used the standard software package available for TIP users to remove the bias level, apply the flatfield correction, 
and demodulate the data to obtain the four Stokes parameters. The flatfield correction 
removed most of the polarized interference fringes, leaving a residual pattern with amplitudes smaller 
than 10$^{-4}$ I$_\mathrm{max}$. Despite the good seeing conditions and the image stability offered 
by the AO system, some occasional image motion was unavoidable within 
the $250$~ms integration time of one modulation, yielding spurious polarization signals (seeing-induced crosstalk).
Normally this effect is minimised by the spatio-temporal demodulation scheme \citep{tip}. 
However, close to the limb, the huge intensity gradients make observations extremely sensitive to this effect. 
As a result, thin stripes of spurious positive and/or negative polarization parallel to the limb
were apparent in the observations. At this point, our observational procedure proved to be very helpful. 
As we observed with the slit always crossing the limb, we were able to 
correct for the cross-talk between Stokes parameters induced by image motion during the integration time 
\citep{yo_12}.

The first and second row of Fig. \ref{intensidad_he} display intensity images of the observed prominence at the core of 
the He\,{\sc i} line with two different saturation levels. The brightest parts of the prominence (the vertical pillars) 
are associated with filament barbs, while the main body of the prominence is fainter and mostly formed by 
thin threads parallel to the surface (see the saturated images of the prominence).
The spatial resolution allowed us to identify 
coherent structures of brightness such as the double-helix seen in both barbs in the third scan, as 
well as the thin filaments forming the main body of the prominence. The helical appearence of the 
prominence legs, made us identify the target with a tornado prominence, as defined by \cite{pettit_32}. 

The remaining rows of Fig. \ref{intensidad_he} display
the amplitudes of Stokes parameters in the He \textsc{i} line. The Stokes $Q$ amplitudes are positive (the reference of Stokes $Q$ is set parallel to
the local limb) in the full prominence, indicating that the anisotropic radiation was mainly coming from below. 
The Stokes $U$ and $V$ amplitudes are weaker than Stokes $Q$ and are induced by the presence of a magnetic field. All Stokes parameters display conspicuous spatial variations mainly in the legs of the prominence. 

Simultaneous images of the tornado prominence were taken with a narrow-band Lyot filter 
centered at the core of the H$_\alpha$ line with a cadence of 1 s and an integration time of 400 ms. 
These images were treated with blind deconvolution techniques \citep{vannoort05} to resolve very 
fine spatial details of the temporal evolution of the prominence. These observations started with the first spectro-polarimetric scan and lasted until the last one was finished. An animation of the full H$_\alpha$ time series is available in the accompanying online material 
of \cite{yo_15}. The H$_\alpha$ data revealed that the motion of the fibrils of the 
observed prominence was sinusoidal \citep[see Fig. 5 in][]{yo_15}, the same as observed in AIA 17.1 nm by \cite{su_12}, who identified 
their observed structures as Pettit's tornadoes. This, together with the helicoidal shape of the prominence, made us identify the legs of our observed prominence with a tornado phenomenon. 

To strengthen such identification, we looked at the AIA 17.1 nm data at the time 
of the observations. It turned out to be very difficult to identify the SDO/AIA 17.1 nm tornadoes at the very time of 
the spectro-polarimetric observations because of the strong glow produced by scattering at the limb. However, using an unsharp masking technique, we were able to identify the structures as soon as 5 h after the spectro-polarimetric observations (see left column in Fig. \ref{images_aia}). To guide the reader, the squares in these panels identify the 
region of the spectro-polarimetric scan. The work of \cite{wedemeyer_13} shows that the average lifetime of solar tornadoes is about 35 h. Therefore, we can be confident that there is a 
direct relationship between our spectropolarimetric results and the 
properties of the tornado that is observed later on with AIA.

Figure \ref{images_aia} shows the temporal evolution of the tornado prominence since it appeared at the East limb 
until it suddenly erupted and disappeared 53 h later. We identify the dark structures in the 17.1 nm images with the solar tornadoes 
as defined by \cite{su_12} and \cite{wedemeyer_13}. In most of the frames, the tornadoes seem 
to have a helicoidal shape, which is better visible at the moment of the eruption when the field 
lines untwist (assuming the brightenings 
follow field lines). As it is rather common in tornado prominences, the main body is fainter and less dense 
as compared to the legs that host the tornadoes.

\begin{figure}[!ht]
\center
\includegraphics[width=\columnwidth]{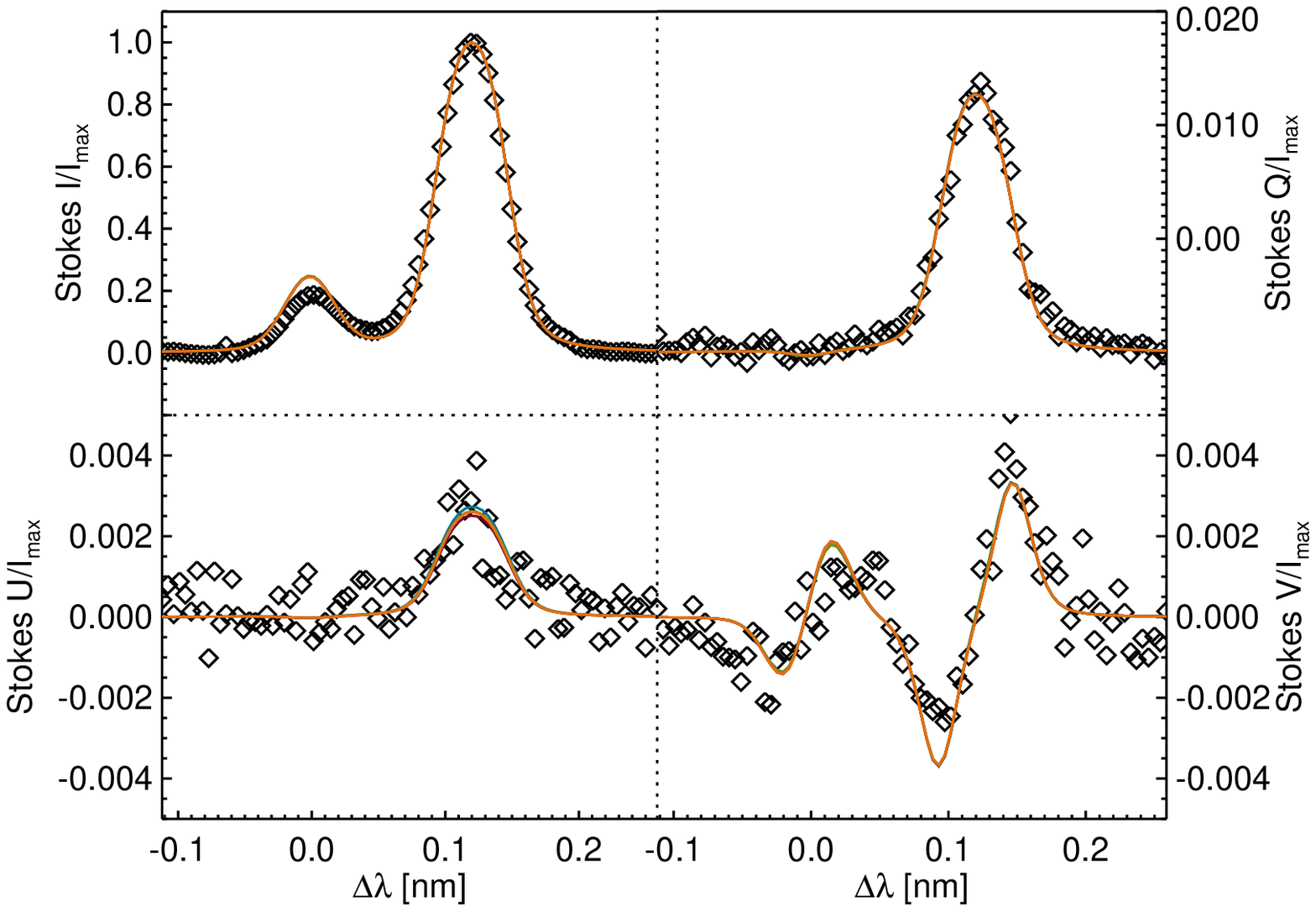}
\includegraphics[width=\columnwidth]{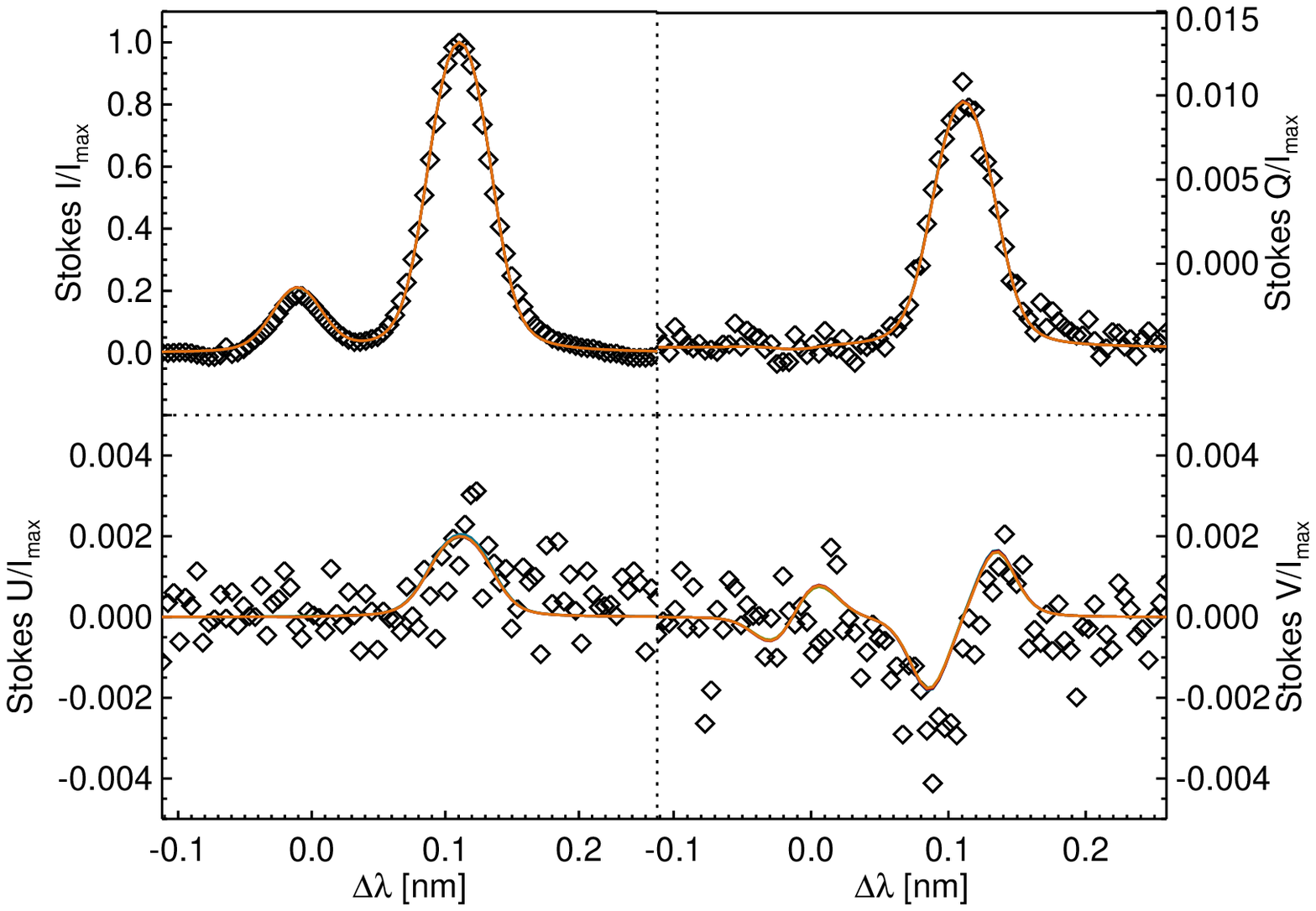}
\includegraphics[width=\columnwidth]{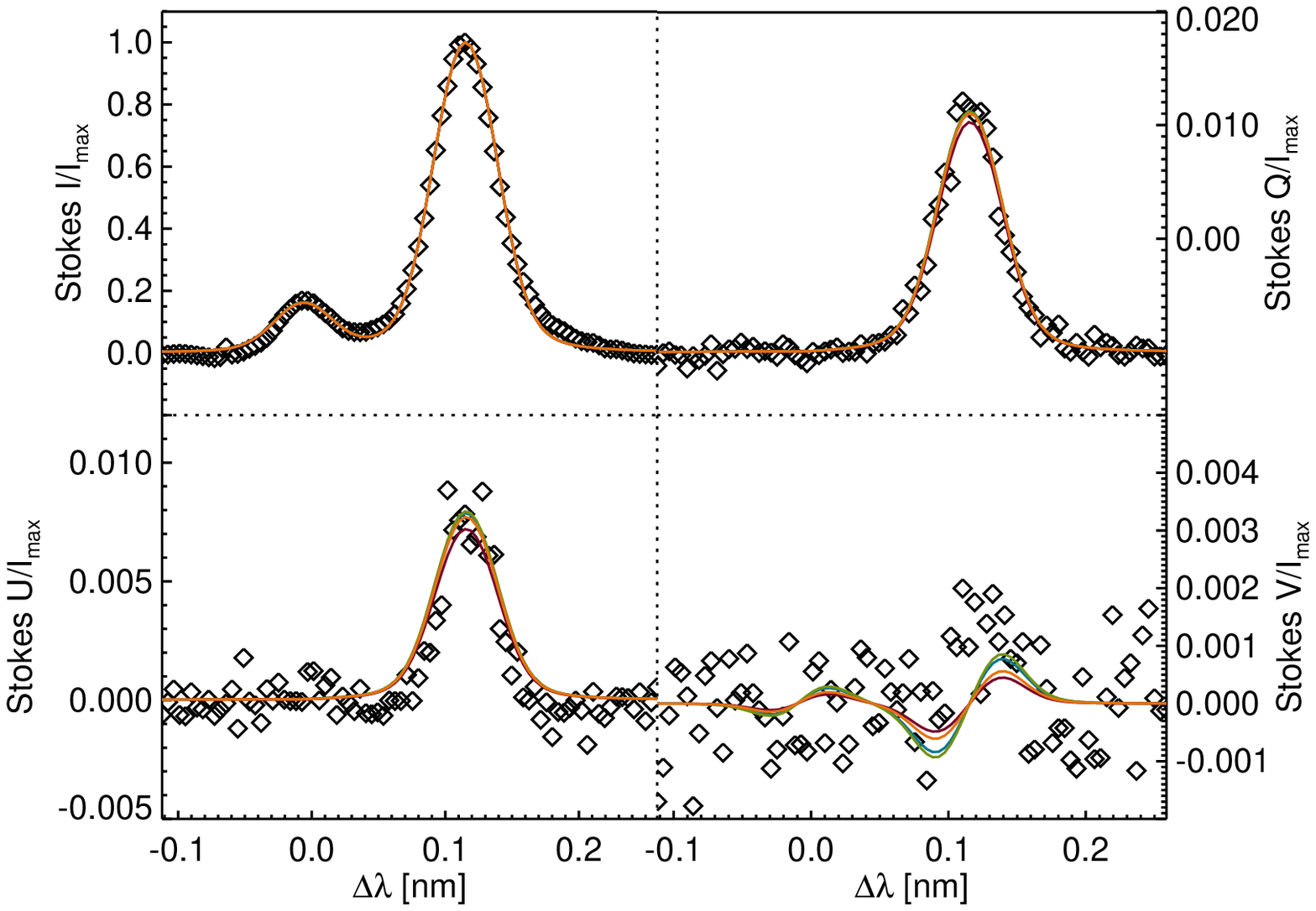}
\caption{Examples of the Stokes profiles recorded at three pixels. The four ambiguous solutions --two horizontal 
and two vertical --found 
with HAZEL are overplotted with different colors. In most cases, the four lines are overlapped and only one color is 
visible.}
\label{examples}
\end{figure}

\begin{figure*}[!t]
\center
\includegraphics[width=\textwidth, bb= 35 21 444 94]{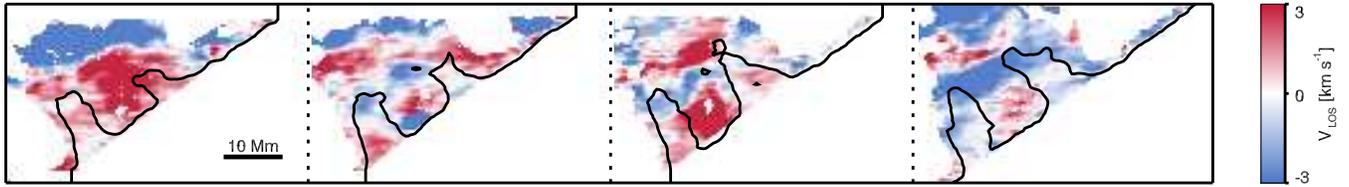}
\caption{Maps of Doppler velocity of the observed prominence 
as inferred from the intensity profiles of the He\,{\sc i} line 
using the inversion code HAZEL. }
\label{vel}
\end{figure*}

\begin{figure}[!t]
\center
\includegraphics[width=\columnwidth, bb= 50 9 450 206]{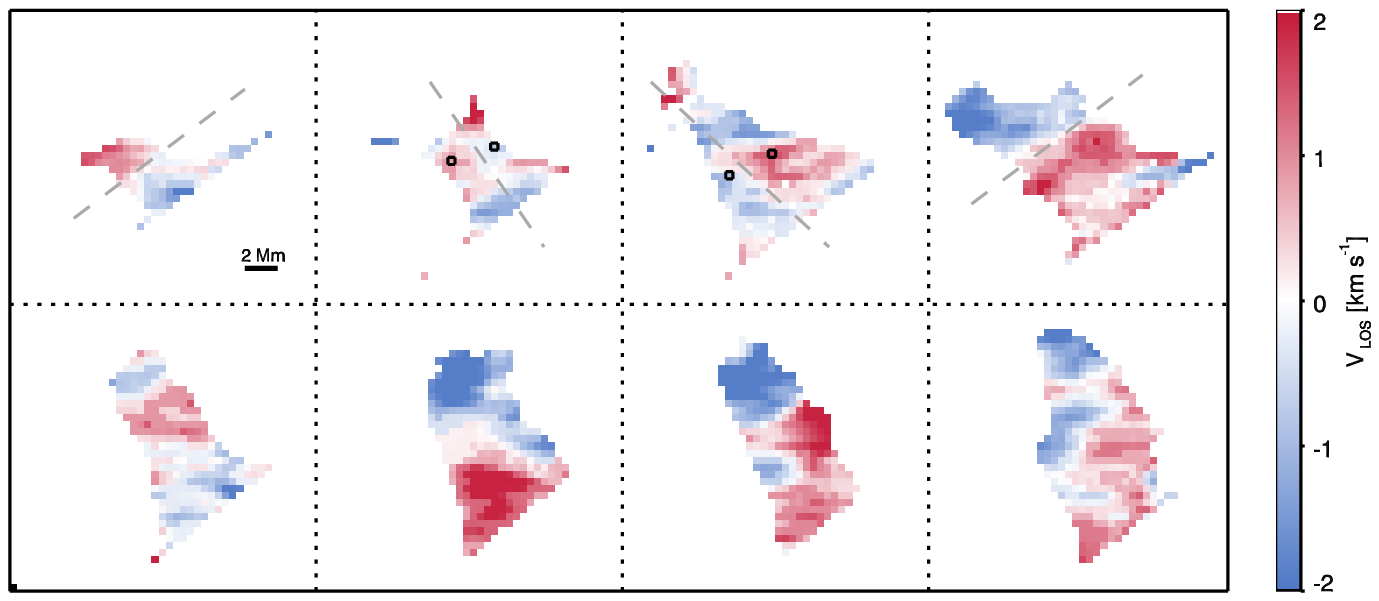}\\
\includegraphics[width=\columnwidth, bb= 50 20 550 220]{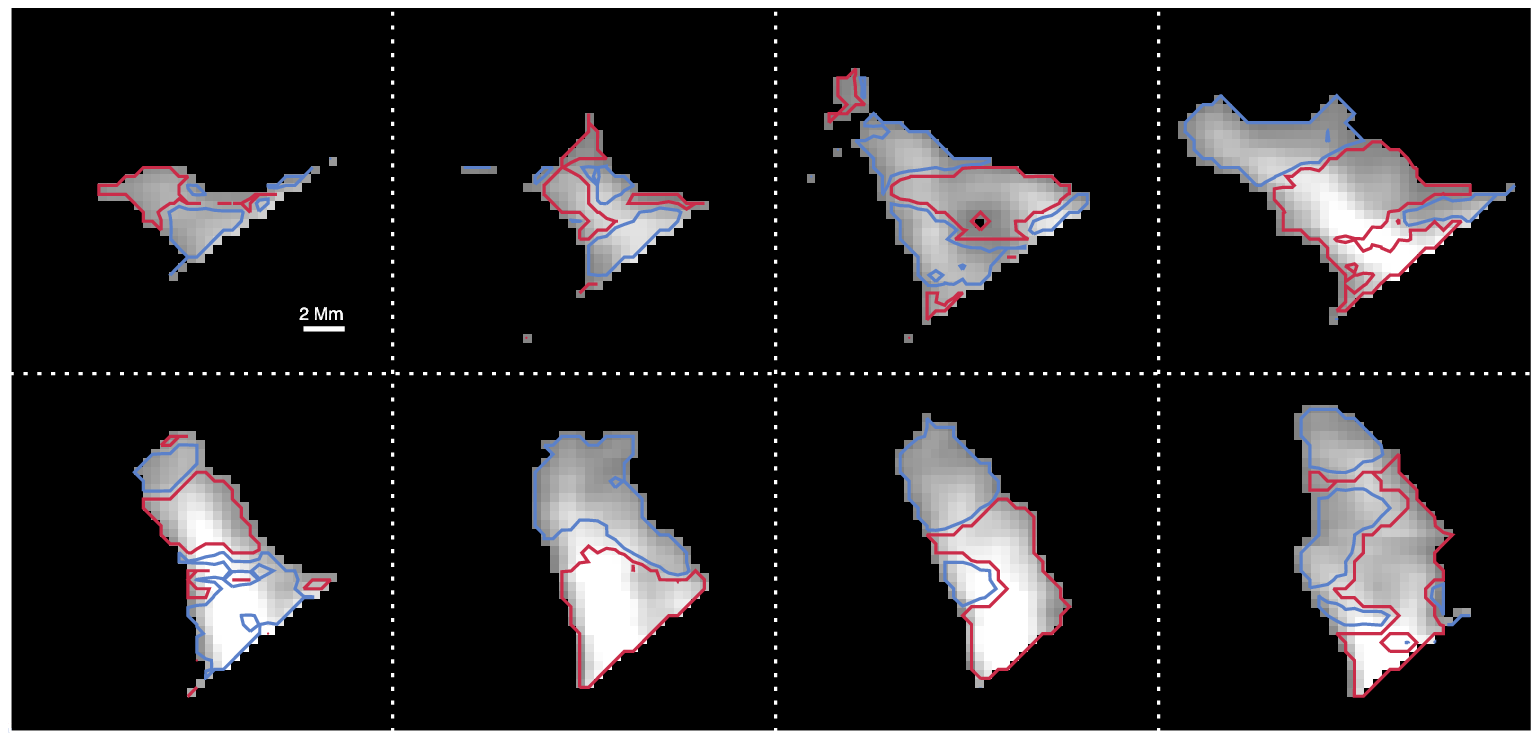}
\caption{Top display: evolution of the Doppler velocities at the legs of the prominence. The velocity is relative to each structure, i.e., it has been computed by substracting the average velocity at each leg. Bottom display: evolution of the intensity (background image) and the Doppler velocity (contours at $\pm 0.2$ km s$^{-1}$) at the legs of the prominence. In both displays the rightmost (leftmost) leg is plotted in top (bottom) panels.}
\label{vel_pies}
\end{figure}

\subsection{Spectropolarimetric inversions}
We used the HAZEL inversion code \citep{hazel} to infer the thermodynamical and magnetic properties of our 
observed prominence. The code solves the forward problem by integrating the radiative transfer equation for polarized light 
on a slab of constant physical properties. It consistently includes scattering polarization and the effect of a 
magnetic field in the combined Hanle and 
Zeeman regimes. The slab is characterized by its height above the surface that controls the
anisotropy of the radiation field (we fix its value 
from the observations), its optical depth, its bulk velocity, 
and its Doppler broadening. The external magnetic field is characterized by the 
field strength and the inclination and azimuth in the local reference frame. There can be up to eight 
ambiguous solutions for the magnetic field vector that yield nearly the same emergent Stokes 
parameters. The number of these ambiguous solutions depends on the actual values of the spectro-polarimetric 
signals, and it can be reduced to only four when the Stokes $V$ signal is above the noise level \citep[for more details see][]{casini_05,judge_07,yo_15}.

For a reliable characterisation of the magnetic field, it is crucial to correctly set the scattering
geometry of the observations. To this aim, we obtained the longitude of the observed prominence from the SDO/AIA 17.1 nm images. This
fixes the scattering geometry and the height of the prominence in each spectropolarimetric scan \citep[for more details see][]{yo_15}. 
Following the standard procedure with HAZEL, the inversion is performed in two steps. First, we inverted the intensity to infer 
the thermodynamical quantities. This information was then fixed in the second step, in which we included the polarimetric information to 
obtain the magnetic field vector. 

Figure \ref{examples} displays three characteristic examples of observed profiles (diamond symbols). The Stokes profiles
synthesised with the inferred parameters are displayed by solid lines. Each color corresponds to a different
ambiguous solution. The Stokes $V$ signal is well above the noise in the first example, 
decreases in the middle panel and is close to the noise level in the bottom one (although the polarity can
still be detected). For this reason, just four ambiguous solutions were found.

A single magnetic atmosphere based on a simplified slab model fits quite well the four Stokes parameters in the
majority of the field of view. However, only in a few pixels we found clear hints of the presence of another 
magnetic component or even gradients along the line-of-sight (LOS), pointing out the presence of radiative 
transfer effects, e.g., the asymmetric Stokes $V$ profile in the first and second examples or the different wavelength shift or broadening
between Stokes $U$ and Stokes $Q$ in the third example.

\section{Results}

\subsection{Dynamics of the prominence}

We inferred the LOS Doppler velocity using only the information of Stokes $I$. We transformed it to absolute velocities
by compensating for the solar rotation, the relative velocity between the Sun and the Earth, and the gravitational shift \citep{kuckein_v12}. This correction has to be done at each slit position because the effects are time-dependent. For simplicity, 
we assumed that all the observed pixels record information coming from the same solar longitude. 
The ensuing velocity maps are shown in Fig. \ref{vel}. 
The prominence shows large velocity patches towards and away from the observer. The prominence body material (in our case, the most tenuous part 
of the observed intensities), in general has a positive-negative velocity pattern parallel to the surface, which is more 
evident in the third and fourth scans. The legs have a positive-negative velocity pattern at both sides of the structures, i.e., almost perpendicular to the limb, during the sencond and third scan. 
On the contrary, this pattern looks parallel (very similar to that of the main body) during the first and last scan. The 
Doppler patterns are similar in both legs, in particular, in the lower heights since higher in the legs Doppler velocities are 
more contaminated by the velocities of the main body.

Assuming that all of the structure is formed at the same longitude is a first-order approximation. In Fig. \ref{images_aia}, we can see that different 
legs of the prominence are located at different solar longitudes and can have different velocities due to the rotation (and, potentially, due to flows). To study in more detail the dynamics of prominence legs, we subtracted the average velocity in each leg from the Doppler velocities. 
With this, we obtain velocities relative to the structure (see top display of Fig. \ref{vel_pies}). The Doppler patterns described before become now more evident: In the first and last scans, an opposite-sign velocity pattern is observed parallel to the limb (we have plotted dashed lines at the zero velocity axis). In the second and third scan this pattern is almost perpendicular to the limb, i.e., opposite Doppler shifts are found at both sides if promience legs. The third scan of the rightmost leg shows that the opposite sign velocities correspond to the two fibrils that form the leg (see bottom display of Fig. \ref{vel_pies}). In the last scan, though, the opposite-sign velocity pattern is clearly parallel to the limb in the rightmost leg, but it correlates with the two fibrils in the leftmost leg (similar to the third scan of the rightmost leg).

\begin{figure}[!t]
\center
\includegraphics[width=0.5\columnwidth, bb = 20 5 265 333]{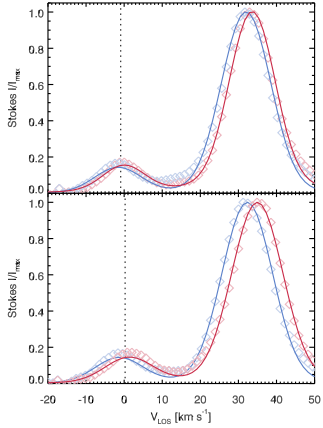}
\caption{Intensity profiles selected from the oppositely-signed Doppler velocities at the borders of the rightmost leg of the prominence (the positions are marked with squares in Fig. \ref{vel_pies}). Top (bottom) panel displays the profiles from the second (third) scan. Diamonds represent the observed profiles and the continuum lines display their best fit. Red (blue) colors are used for those profiles that are red (blue) shifted with respect to the average Doppler shift at each leg (marked as a vertical dashed line).}
\label{vel_rel}
\end{figure}

The differential velocities between the red and the blue-shifted patches are often below 2 km s$^{-1}$. In the second and third scan, the rightmost leg has the smallest velocities but they still have amplitudes that are several times the velocity resolution of the data ($\sim 0.12$ km s$^{-1}$). Figure \ref{vel_rel} displays typical profiles of the red and blue-shifted areas of the rightmost leg during the second (top panel) and the third (bottom panel) scan. As can be seen, the profiles are clearly blue- and red-shifted with respect to the average velocity in the leg.

\subsubsection{Rotation scenario}

Positive and negative Doppler velocities at both sides of the legs of solar prominences, like the ones
we observe in the second and third scans of Fig. \ref{vel_pies}, have been considered as an observational signature of rotational motions 
\citep{david_12, wedemeyer_13,su_14,levens_15}. In the prominence under study, the simultaneous H$_\alpha$ line-core photometry shows that the 
fibrils of the legs follow a periodic motion, with a period of $\sim$50-60 min \citep[see Fig. 5b in][]{yo_15}. This
kind of photometric patterns have also been interpreted as rotation \citep{su_12}. All in all, our observations, however, 
do not fully support the rotation scenario for the following reasons:

\begin{itemize}
\item With a rotation period close to 60 min as inferred from the photometry, the periodic motions should
have tangential velocities of up to $\sim$9 km s$^{-1}$. This is much larger (by more than a factor 4) than the velocities inferred
from the Doppler effect in the He\,\textsc{i} line (see Fig. \ref{vel_pies}). However, one must always be
careful to directly interpret changes in H$_\alpha$ as real plasma motions because they can also be due 
to changes in the thermal conditions.

\item The positive-negative velocity pattern at both sides of the prominence legs reverses between the second and the third scan. 
This means that a sustained rotation --if it exists-- cannot last for more than $\sim$60 min. Given the size of the prominence legs ($\sim 4.5$ Mm) and the inferred LOS Doppler velocities (less than 2 km s$^{-1}$), the prominence should take around 4 h to make a complete turn.
Note also that a similar behavior is found in Fig. 5 of \cite{david_12}, although they explained their observations
as a swirling motion.

\item The positive-negative velocity pattern at both sides of the tornadoes is not detected in the first and 
last scan. This means that the rotation -- if it exists at that time-- is inhibited.

\end{itemize}

Our observations are strongly in contrast with the ones presented by \cite{su_14} that show a coherent Doppler pattern 
of the structure --which the authors interpret as rotation-- for more than 3 h. They observe the extreme ultraviolet 
Fe\,{\sc xii} line and deduce Doppler velocities that are very close to the sensitivity of the EIS instrument 
\citep{eis}. We note that their figures show a positive-negative velocity pattern not only in the 
tornado, but also in the surrounding area, which might indicate the presence of a systematic effect \citep{levens_15}.
Compensating for the systematic effects could make their and our observations compatible. Yet, another 
possible explanation is that either their object or ours are particular cases of tornadoes. Though this seems 
unlikely, we can not rule out this possibility.

\begin{figure}[!t]
\center
\includegraphics[width=\columnwidth, bb = 61 37 486 250]{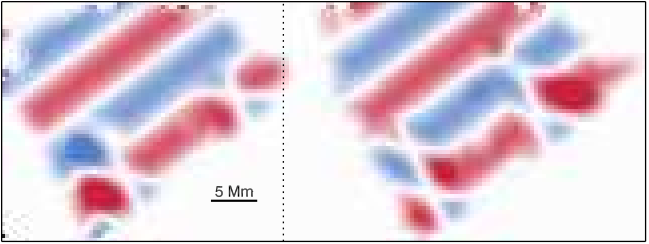}
\caption{Numerical simulation of a superposition of two sinusoidal perturbations propagatins as waves, one 
across the full prominence and another one restricted to the legs. The quantity plotted is the Doppler 
velocity relative to its maximum amplitude, hence without units (i.e., the quantity in Eq. 1).}
\label{simu_osc}
\end{figure}

\subsubsection{Wave scenario}

An alternative explanation for our measured Doppler shifts is the presence of a periodic disturbance of the prominence plasma, 
which might be propagating as a wave. Waves and oscillations are commonly observed in solar prominences and are classified 
in terms of small/large amplitude oscillations \citep[see][for reviews]{oliver_02,arregui_12}. 
Their presence was invoked by \cite{panasenco_14} in explaining the sinusoidal motion of the fibrils of prominence legs associated to tornadoes observed by SDO/AIA in the 17.1 nm channel. The inferred oscillation periods were in the range of 40-70 min, depending on the height above the surface. More recently, \cite{mghebrishvili_15} also reported the detection of quasi-periodic transverse displacements, interpreted as MHD kink waves, in a solar prominence tornado.  Their time distance analyzes showed two patterns of quasi-periodic brightness displacements with periods of about 40 and 50 min. 

Of particular interest to our study is the analysis performed by \cite{terradas_02} on the spatial distribution of Doppler velocity oscillations in the body of a quiescent  prominence, using time series of two-dimensional H$_\beta$ Dopplergrams. Their temporal and spatial analysis enabled them to detect the simultaneous presence of waves with standing and propagating character, to analyze their damping properties, and to quantify wave properties such as the direction of propagation, wavelength, and phase speed. The reported periods are of about 70-80 min in the upper part of the prominence body and 30-40 min at lower heights, with the waves propagating possibly through a guide. Their results point to a coherence of periods over large patches, with some regions in which positive and negative velocity patters alternate in time. The spatial coherency of their signals and the values of the reported periods indicate that they are of solar origin and not of instrumental or seeing effects \citep{zapior_15}.

Because of the lack of an appropriate temporal cadence in our data, a detailed analysis akin to that of \cite{terradas_02} cannot be applied here. We can, however, follow a simple, qualitative procedure to test the compatibility of the wave scenario against the observed data.
Let us assume that whatever the nature of the perturbation is that causes the observed Doppler velocities, it can be represented as a periodic perturbation with a direction of propagation along the vector $\mathbf{k}$. If the $x$- axis points along the slit and that the $y$- axis along the scan direction, the measured perturbation in the Doppler velocity $v_{LOS}$ can be expressed as
\begin{equation}
v_{LOS}(x,y) \propto \int_0^{\Delta t} \mathrm{dt} \cos [k_x\,x + k_y\,y - \omega t + \phi],
\end{equation}
where we have taken into account that a temporal integration is performed for each slit position. By so doing, we are 
simulating the observations, including the time integration per slit position and the slit motion.
The integration time is given by $\Delta t$, $\omega$ is the frequency of the oscillation and $\phi$ is a phase. 

By way of example, Figure \ref{simu_osc} shows simulated velocity patterns obtained following this procedure, corresponding to our second and third scans. Doppler patterns are visible both in the main body of the prominence and in the legs showing different 
behaviors: the main body has a pattern mostly parallel to the local limb while the legs have a mostly perpendicular-to-the-limb Doppler pattern. Then, to simulate the observed Doppler pattern, two oscillation components within two different waveguides (the main body and the legs) were linearly combined. The first one, applied to the main body of the prominence, consisted of a perturbation propagating perpendicular to the local limb with a period of 80 min and a wave vector with magnitude $k = 2\pi/15$ arcsec$^{-1}$. Note that, due to the solar curvature, the main body is seen projected down to the limb. The second component had a period of 50 min and a wave vector of magnitude $k = 2\pi/9$ arcsec$^{-1}$ and was applied only to the prominence legs. It formed an angle of 24$^\circ$ with respect to the local limb. In our simulated patterns, period and wavelength values were chosen to qualitatively fit the observed velocity pattern that is mostly parallel to the surface.  Also, we applied phase shifts of 200$^\circ$ and 70$^\circ$ to the leftmost and rightmost legs, respectively. Since we are not interested in amplitudes but in Doppler patterns, we plot the velocity relative to its maximum amplitude, hence 
without units.

Our example simulation constitutes a particular solution that is able to reproduce qualitatively the observed velocity magnitudes and pattern,   including the alternate positive and negative evolution of the velocities at the legs of prominences with time. Furthermore, the necessary periods and wavelengths are in good agreement, in order of magnitude, with those reported by \cite{terradas_02} and \cite{panasenco_14} .

\begin{figure*}[!t]
\center
\includegraphics[width=\textwidth,bb=36 17 448 120]{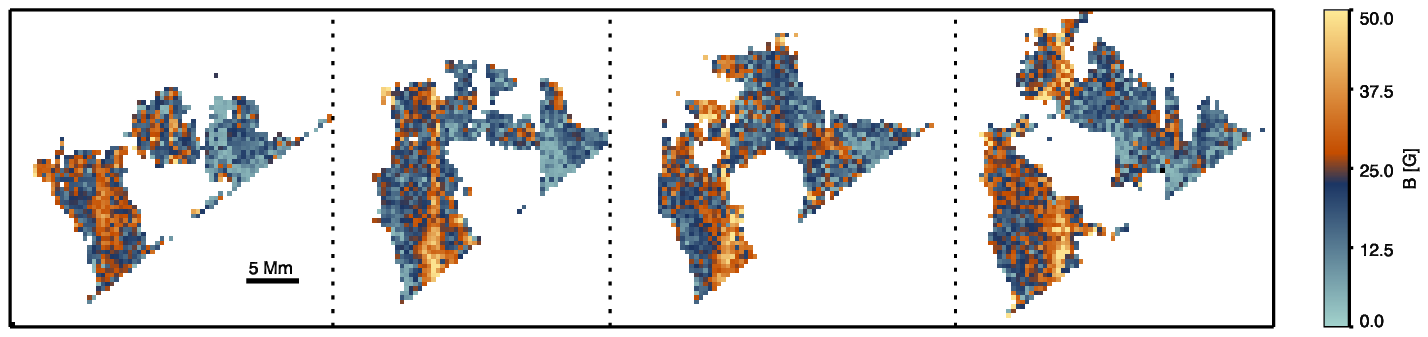}\\
\includegraphics[width=\textwidth,bb=36 21 550 141]{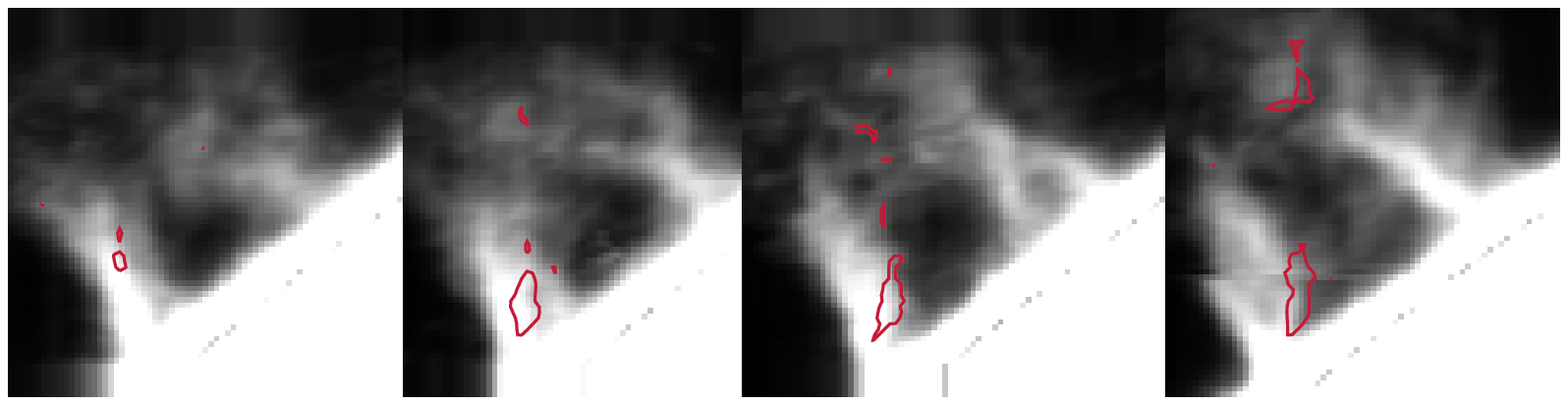}
\caption{Magnetic field strength of the prominence inferred from the spectro-polarimetric data 
of the He\,{\sc i} line. The bottom panels display the line-core intensity images with an isocontour of the 
magnetic field strength of 32 G.}
\label{b+contour}
\end{figure*}

\begin{figure*}[!t]
\center
\includegraphics[width=\textwidth,bb=36 22 445 210]{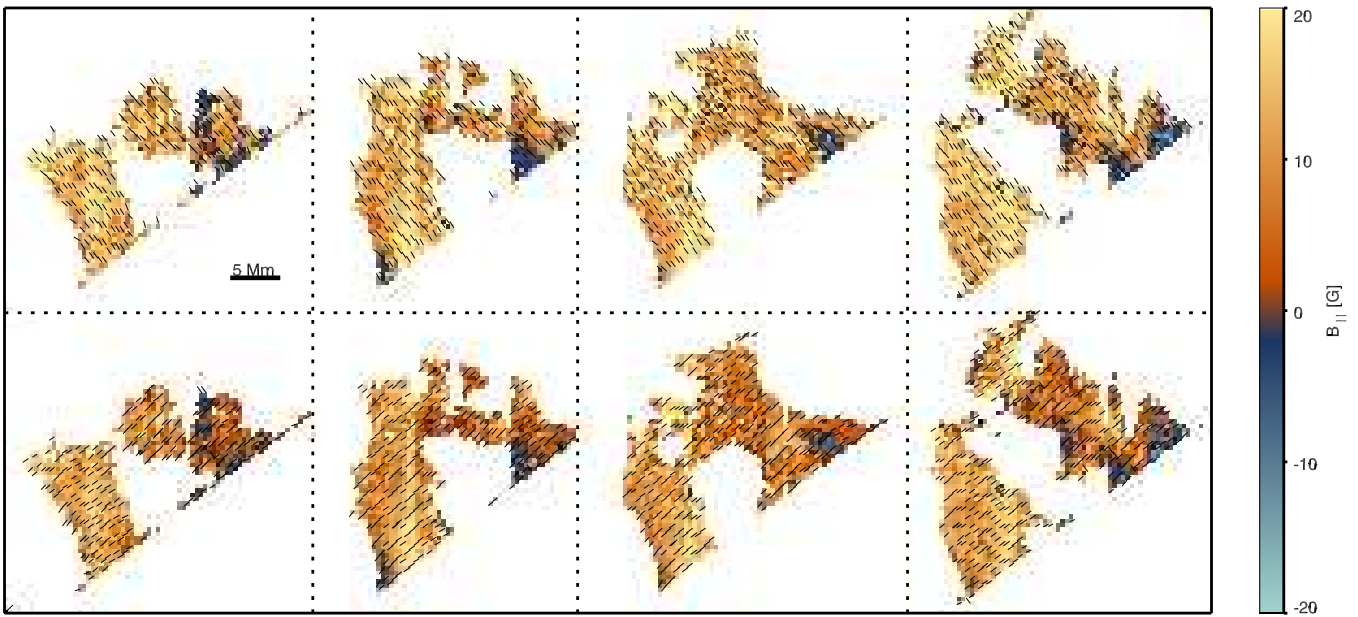}
\caption{Inferred topology of the magnetic field at the prominence. The background images 
display the longitudinal magnetic field and the lines represent the direction of the 
projection of the magnetic field onto the plane of the sky. Top (bottom) panels correspond to the family of solutions with vertical (horizontal) magnetic field. The two solutions within a family will have a 180-deg difference in the plane-of-the-sky projected magnetic field.}
\label{blos+proyeccion}
\end{figure*}

\begin{figure*}[!t]
\center
\includegraphics[width=\textwidth,bb=49 31 490 243]{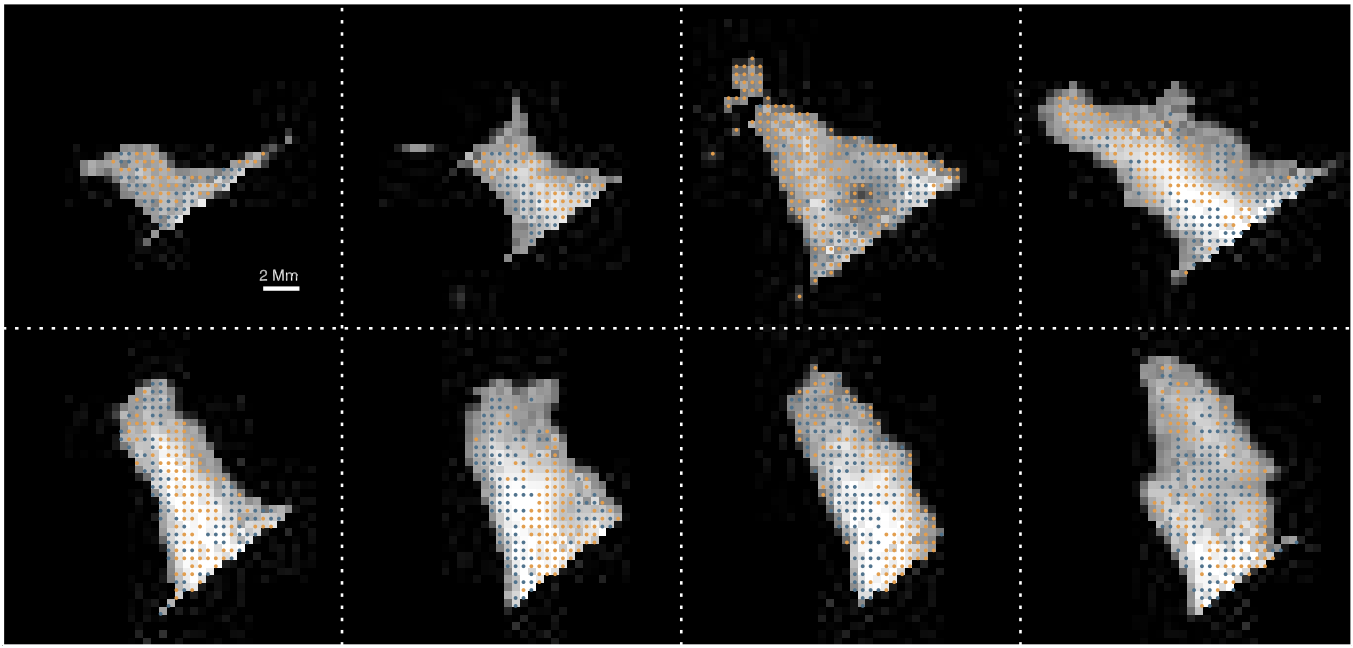}
\caption{Evolution of the intensity (background image) and the sign of the LOS component of the magnetic field at the legs of the prominence. Blue (orange) represent the negative (positive) values of the LOS longitudinal field component. The rightmost (leftmost) leg is plotted in top (bottom) panels.}
\label{mapai+bpar}
\end{figure*}

\subsection{Magnetic field of the prominence}

As explained before, four ambiguous solutions exist at each pixel, which would potentially lead to $4^n$ (with $n$ the number of
pixels) magnetic field configurations that are compatible with the observed Stokes profiles.
We solved this combinatorial problem by imposing the additional constraint of minimizing pixel-to-pixel gradients.
This led to four ambiguous magnetic field configuration for the whole structure that could be grouped into two families: 
those with fields preferentially perpendicular or parallel to the local solar limb. The two members of each family display a projection
of the magnetic field in the plane of the sky that differs by 180$^\circ$.

The determination of the magnetic field strength is very robust, having similar values in all ambiguous solutions, with a median
difference of 3 G. The top panel of Fig. \ref{b+contour} displays the magnetic field 
strength in all pixels with polarimetric signals above three times the noise level. Most of the fields are below 20 G but there 
are regions where the magnetic field increases to 40-60 G, coinciding all the time with the border of a filamentary vertical structure at 
the leftmost legs of the prominence (see bottom panels of Fig. \ref{b+contour}). We speculate that the fibril is the 
same in all four scans and it slightly moves from up-left to down-right. If this motion is real, the fibril moves at 0.6-1 km s$^{-1}$, 
a velocity that is compatible with the observed Doppler shifts. However, it is much smaller than the plane-of-the-sky apparent motions of the H$_\mathrm{\alpha}$ line \citep[$\sim 26 $ km s$^{-1}$; see Fig. 5 in ][]{yo_15}. However, it could also be that the physical properties are changing with time, making the motion of the fibril apparent. 

As can be seen in Figure \ref{blos+proyeccion}, in both ambiguous solutions the projection of the field onto the plane of the sky is at an angle 
with the axis of each fibril that form the prominence legs. We also can observe a reversal of the polarity of the LOS magnetic 
field at opposite sides of the fibrils that form the legs (see Fig. \ref{mapai+bpar}). With these findings, we infer helical fields 
in the fibrils of the legs of the prominence. The rightmost leg is splitted in two fibrils in the last scan. It can be seen that 
the rightmost fibril has a negative-positive pattern from left to right, the same observed in the fibrils of scans one to three. The same 
behavior can be observed in the leftmost leg. In this case, is is splitted into two fibrils in scan three. We 
again speculate that we are facing the motion of the fibrils of the prominence and not changes in the physical properties with time. 

Overcoming the intrinsic Hanle ambiguities spectroscopically is only potentially possible by the simultaneous 
observation of many spectral lines \citep{egidio_bommier93,casini09}. In \cite{yo_15} we could disambiguate the problem using 
additional physical constraints, but only in the legs of the prominence. We use the same method to study the time evolution 
of the topology of the magnetic field in the legs of the prominence. We invoke the stability of the structure against a 
kink perturbation since it still lives for more than two days. The Kruskal-Shafranov criterion states that a kink stability develops 
when the twist of the magnetic field exceeds a critical value, so that a structure is stable if 
\begin{equation}
\frac{2\pi r}{L}\frac{B_z}{B\theta} \ge 1.
\end{equation}
The symbols $r$ and $L$ are the radius and the length of the fibrils, respectively, and $B_z$ and $B_\theta$ are the vertical and azimuthal components of the magnetic field in polar coordinates. From the intensity images, we estimate $r=1.7$ Mm and $L=11.6$ Mm. 
The stability criterion has values between 0.1-0.2 for the horizontal solution and between 1.3-1.9 for the vertical solution. 
Since the prominence still lives for more than two days, we deduce that the fibrils of the prominence feet have vertical helical 
magnetic fields in all sequences. Unfortunately, for the main body of the prominence we can not disambiguate the problem.

\section{Discussion}

The magnetic field in the legs of our observed prominence is helicoidal, connecting the main body with the underlying surface. 
The helical field is rather stable during the whole observation period, of about two hours. We observe 
a slow temporal evolution of the topology (and strength) of the magnetic field in the legs of the prominence, probably 
due to the proper motion of the magnetic skeleton of the fibrils that form the legs, rather than a change in the plasma properties that could modify the properties of the magnetic field. 

Fibrils that move with respect to each other have been already reported and interpreted as rotation around a common 
axis \citep{mghebrishvili_15}. In our case, this is unlikely because the structures would take several hours to perform a turn, while the Doppler patterns change in less than one hour. From the observational data presented in this work, any type of rotation --if existent-- must be intermittent on a time scale of less than one hour. 

We propose that the connectivity of the prominence spine,
with a well-defined helicity, and the photospheric magnetic
field below, with a fluctuating topology, may naturally yield the
kind of helical structures we find and their temporal variation. Changes in the photospheric magnetic
field can affect the topology of the prominence legs \citep{wedemeyer_12} and hence its twist. For example, twisting and untwisting field lines can appear as opposite rotation motions of fibrils of prominence legs. Together with upflows or downflows 
along helical lines \citep[as proposed in ][]{yo_15} it can explain the variety of Doppler shifts observed in the legs of the prominence reported in this work. We can not exclude that magnetohydrodynamical waves are also 
responsible for these patters since they are natural to prominences \citep{oliver_02,arregui_12} and can qualitatively explain our observables. A full interpretation in terms of waves requires a proper knowledge of the underlying field structure.

In our opinion, our observed prominence hosting tornadoes in its legs is not a special case of tornado prominence, hence, 
the dynamical and magnetic properties obtained in this work could be extrapolated to other tornadoes. However, we must be 
careful and we encourage for more analysis of spectro-polarimetric observations of solar tornadoes to really assert their dynamics and magnetic field properties.

\begin{acknowledgements}
This work is based on observations made with the German Vacuum Tower Telescope (VTT) and the Solar Dynamics Observatory spacecraft. The VTT is 
operated on Tenerife by the Kiepenheuer-Institut f\"ur Sonnenphysik in the Spanish Observatorio del Teide of the 
Instituto de Astrof\'\i sica de Canarias. The AIA/SDO data is courtesy from NASA/SDO and the AIA science
team. Financial support by the Spanish Ministry of Economy and Competitiveness and the 
European FEDER Fund through projects AYA2010-18029, AYA2014-55456-P, and AYA2014-60833-P are 
gratefully acknowledged. AAR and IA acknowledge financial support through the Ram\'on y Cajal fellowship. 
\end{acknowledgements}


\end{document}